\documentclass[baaa]{baaa}

\usepackage[pdftex]{hyperref}
\usepackage{subfigure}
\usepackage{natbib}
\usepackage{helvet,soul}
\usepackage[font=small]{caption}

\contriblanguage{1}

\contribtype{1}

\thematicarea{2}

\title{Chemical abundance patterns in Local Group galaxies within cosmological simulations}

\titlerunning{Chemical abundance patterns in simulated galaxies}

\author{
L. Biaus\inst{1},
C. Scannapieco\inst{1}
\&
S.E. Nuza\inst{2, 1}
}

\authorrunning{Biaus et al.}

\contact{lbiaus@df.uba.ar}

\institute{
Departamento de Física, Facultad de Ciencias Exactas y Naturales, UBA, Argentina
\and
Instituto de Astronom\'ia y F\'isica del Espacio, CONICET--UBA, Argentina
}

\resumen{
En el contexto del modelo cosmológico de concordancia, la formación de estructura en el Universo es el resultado de la amplificación, por efectos gravitatorios, de pequeñas perturbaciones en el campo de densidad primigenio. Esto resulta en la formación de estructuras conocidas como halos de materia oscura, donde el gas colapsa y forma estrellas, dando lugar a galaxias. Las simulaciones numéricas son una herramienta importante en el estudio teórico de la formación y evolución de galaxias. En este trabajo describimos la implementación de un modelo de enriquecimiento químico en simulaciones cosmológicas del Grupo Local de última generación. Éstas incluyen modelos de subgrilla para los procesos físicos más relevantes. Analizamos la evolución química y morfológica de dos galaxias con masas viriales similares a nuestra Vía Láctea. Para cada una de las componentes estelares (disco, bulbo y halo), establecemos conexiones entre su historia de formación y su evolución química. Encontramos que el enriquecimiento de elementos $\alpha$ (O, Mg, Si) ocurre en etapas tempranas de la evolución, pues sus principales productoras son estrellas de vida corta que devienen en explosiones de supernova tipo II. Hay también una contaminación gradual del resto de los elementos a medida que ocurren supernovas de tipo Ia y vientos de estrellas en la fase asintótica gigante.
}

\abstract{
In the context of the concordance cosmology, structure formation in the Universe is the result of the amplification, by gravitational effects, of small perturbations in the primeval density field. This results in the formation of structures known as dark matter haloes, where gas collapses and forms stars, giving birth to galaxies. Numerical simulations are an important tool in the theoretical study of galaxy formation and evolution. In the present work, we describe the implementation of a chemical enrichment model in a  state-of-the-art cosmological simulation of the Local Group. The simulation includes sub-grid models for the most relevant physical processes. We analyze the chemical and morphological evolution of two galaxies with virial masses similar to that of our Milky Way. For each of the stellar components (disc, bulge and halo), we establish links between their formation history and their chemical evolution. We find that $\alpha$-element (O, Mg, Si) enrichment happens at early stages of evolution, as their main producers are short-lived stars which end their lives as type II supernova explosions. There is also a gradual contamination with the rest of the elements as type Ia supernovae and winds of stars in the asymptotic giant branch occur.
}

\begin{document}

\maketitle

\section{Introduction}\label{S_intro}

Within the current cosmological paradigm, structure formation occurs in a hierarchical fashion, with smaller systems merging together to form bigger ones. % in a non-linear process.
In this context, numerical simulations are an important tool in the study of galaxy formation, as they can follow the joint evolution of dark matter and baryons in a consistent way, %describing the growth of dark matter haloes and galaxies, and
naturally capturing processes such as mass accretion and mergers. 

During the last years, the possibility of obtaining detailed information on
the chemical properties of stars in the Milky Way, as well as in external
galaxies, opened up an important area in studies of
galaxy formation.
%The relevance of such studies is based
%on the fact that  chemical abundances in the gaseous and stellar components can
%reveal the formation history of a galaxy, in terms of the star formation activity, the time evolution of the insterstellar medium (ISM) and the occurrence of
%events such as mergers.
Chemical elements are synthesized in stellar interiors and ejected into the
interstellar medium (ISM)
in supernova explosions and stellar winds. As stars form, they
inherit  the metallicity of the ISM at that time:
by analysing the chemical abundances of  stars of
different age we can reconstruct the formation history of a galaxy.
On the other hand, by looking at the gaseous abundances of galaxies
at different times, we can infer the properties of the ISM and understand
the effects of the physical processes acting at a given cosmic time.

In this work, we study the chemical properties of a Milky Way-mass galaxy, using a
simulation of the Local Group which is part of the CLUES project. We study the
abundances of various chemical elements at the present time, as well as the
chemical patterns of the different stellar components -- disc, bulge and stellar halo.
This conference proceeding abridged paper is organized as follows. In Sec.~\ref{S_proj} we describe the
simulation, in Sec.~\ref{S_res} we present our results and in Sec.~\ref{S_con} we give our conclusions.

\section{The simulation}\label{S_proj}

\begin{figure}[!t]
\centering
\includegraphics[width=.85\columnwidth]{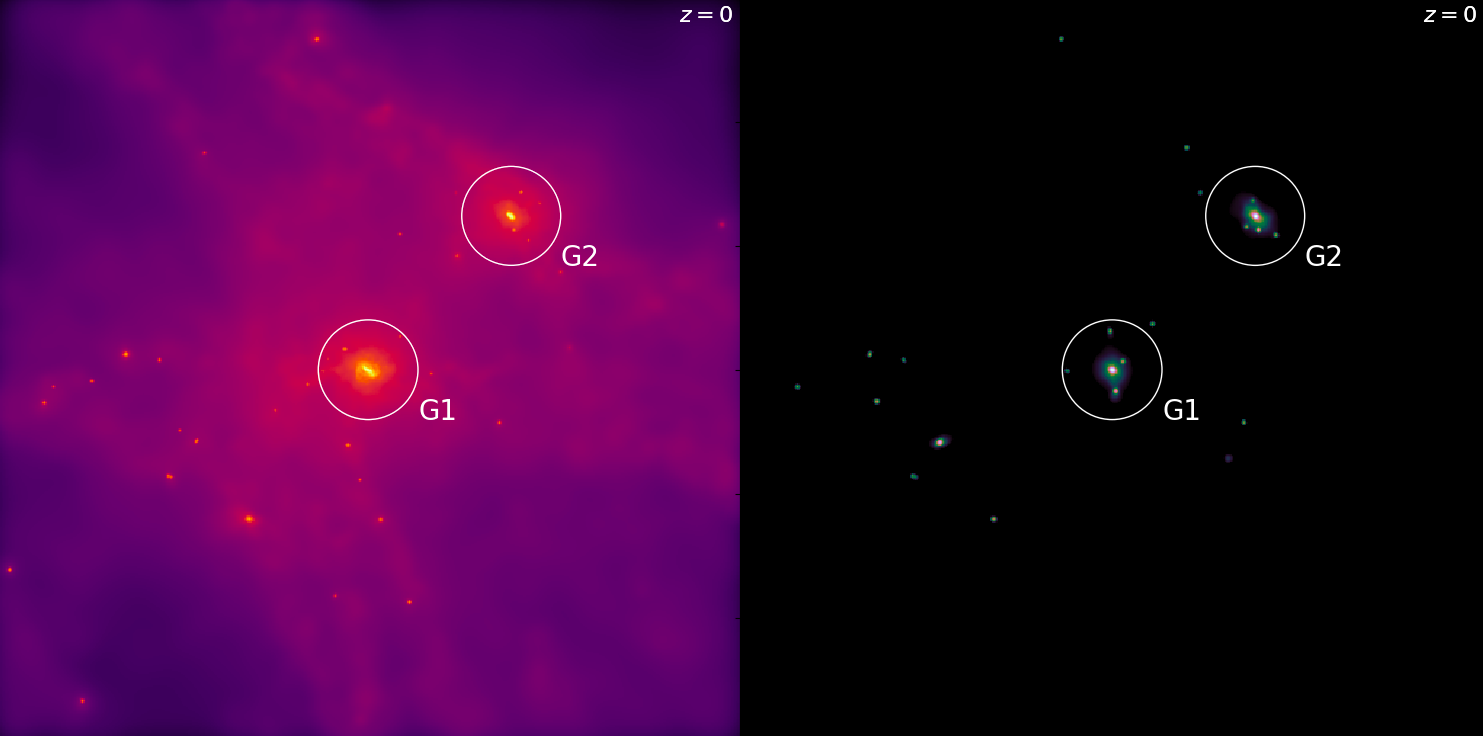}
\caption{\emph{Left panel:} density projection of the gas. \emph{Right panel:} density projection of the stars. Both correspond to the simulated Local Group, for $z=0$.}
\label{fig:proj}
\end{figure}

In this project,  we used  a simulation of the Local Group  that uses initial conditions
of the CLUES project (e.g. \citealt{Scannapieco2015}). The simulation describes the formation of the Local Group, where two Milky Way-sized galaxies form. The galaxies are referred to as G1 and G2 and are, at $z=0$, separated by a distance of $\sim 1~\mathrm{Mpc}$  (see Fig. \ref{fig:proj}). 

The simulation was run using the state-of-the-art code {\sc GADGET 3} \citep{Springel2008} and the chemical enrichment model of \cite{Pierre2018}.
The relevant physical processes which act below the resolution of the simulations are implemented as sub-grid models, namely: star formation, metallicity-dependent gas cooling, supernovae explosions, stellar winds from giant stars and their associated chemical enrichment.
%Initially, the simulations contain a number of dark matter (DM) and gas particles only. Both components obey newtonian gravity which is implemented using a combination of a Tree algorithm and a Particle Mesh method. 
The gas' hydrodynamical equations are solved using the Smoothed Particle Hydrodynamics (SPH) technique.
%Gas particles above a certain density level $\rho_{crit}$ and forming part of a convergent flow ($\nabla \cdot \mathbf{v} < 0$) are stochastically chosen to form star particles. Each star particle represents a stellar population containing approximately $10^5 M_{\odot}$. The initial mass function (IMF) determines the distribution of stellar masses for each stellar population. In this simulation, we adopt the  \cite{Chabrier2003} IMF.

We take into account 3 types of chemical enrichment events: type II and type Ia supernovae (SNe II and SNe Ia respectively), and winds from stars in the Asymptotic Giant Branch (AGB winds). These are all implemented through discrete enrichment events, as follows:

\begin{itemize}
\item SNe II take place shortly after a stellar population's birth ($\tau \sim 10^6~\mathrm{yr}$), with lifetimes depending slightly on metallicity \citep{Portinari1998}. Stars with $M_* \geq 8~\mathrm{M_{\odot}}$ are considered SNe II progenitors. All \item in a star particle are considered to occur simultaneously at the average lifetime of the population.
\item SNe Ia occur once per star particle at a time stochastically determined by a bimodal delayed time distribution \citep{Mannucci2006}.
\item AGB winds gradual enrichment is modelled by 3 discrete events occuring at $10^{-1}$, $1$ and $10~\mathrm{Gyr}$ since the star particle's birth. In these events, stars in the $0.8-7\,$M$_{\odot}$ mass range expel $25\%$, $55\%$ and $20\%$ of their total mass, respectively.
\end{itemize}

%The chemical yields give detail about the chemical composition of the material ejected in the events above. In this simulation,
The yields implemented are mass and metallicity dependant for AGB winds \citep{Marigo2001} and SNe II \citep{Portinari1998}, while for SNe Ia fixed yields for each element were considered \citep{Thielemann2003}. For more details about the chemical enrichment model we refer the reader to \citet{Pierre2018}.

SNe II events also provide an energetic feedback in the form of $10^{51}~\mathrm{ergs}$ distributed among nearby gas particles.

\section{Results}\label{S_res}

In order to identify the different stellar components of the simulated galaxies, we 
 classify stars according to their circularity $\varepsilon$, which is
defined as the ratio between its circular momentum in the $z$-direction and the one corresponding to a circular orbit at that radius ($\varepsilon = \frac{j_z}{j_{circ}}$, where $j_{circ} = r \cdot \sqrt{G M(r) / r}$, $M(r)$ being the enclosed mass up to a certain radius $r$). %Stars in the galaxy disc have a circularity spread centered around unity and stars in the bulge have a spread centered around zero if the bulge is non-rotating. 
Note that, in this work, we only show the  chemical properties of G1, although results are similar for G2.
%We will only show and analyse results and trends corresponding to G1. These are qualitatively similar in G2.

\begin{figure}[!t]
\centering
\includegraphics[width=.5\columnwidth]{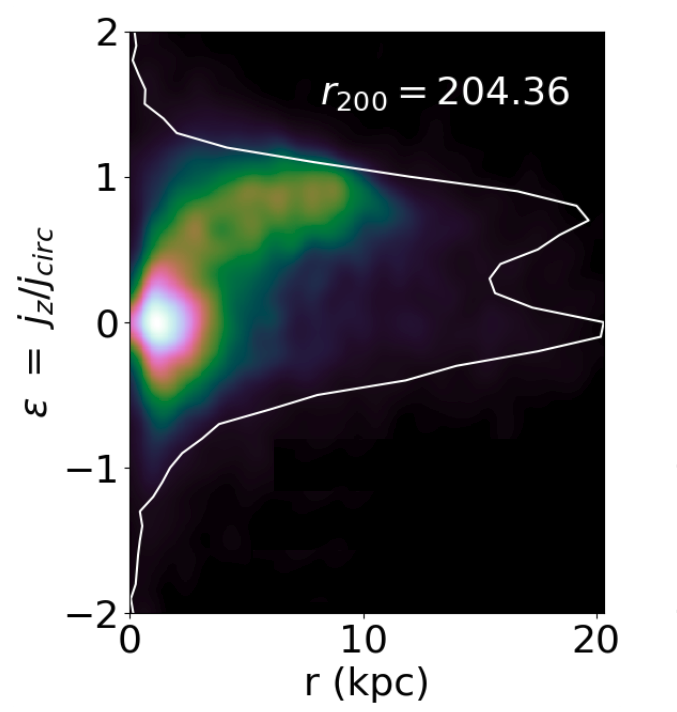}
\caption{Distribution of star particles in the $r-\varepsilon$ plane. The white line shows the circularity distribution added up for all radii. The two distinct peaks correspond to the stellar bulge (centred at $\varepsilon \sim 0$) and the stellar disc (centred at $\varepsilon \sim 1$).}
\label{fig:circ}
\end{figure}

In Fig. \ref{fig:circ} we show the radius-circularity distribution for G1. It can be seen that the bulge dominates the central part of the galaxy ($\varepsilon \sim 0$ at $r < 3~\mathrm{kpc}$) while the disc dominates the outer region up to $r \sim 10~\mathrm{kpc}$ ($\varepsilon \sim 1$). We consider stars with $\varepsilon > 0.5$ to be part of the stellar disc. The remaining ones are considered to be part of the bulge for $r < 10~\mathrm{kpc}$ and part of the stellar halo for $r > 10~\mathrm{kpc}$. %(the bulge is quite smaller, extending up to $r \sim 3$~kpc, but we extend the radial cutoff to avoid contamination in the halo component with bulge stars, as the former has much lower density).

\begin{figure}[!t]
\centering
\includegraphics[width=0.75\columnwidth]{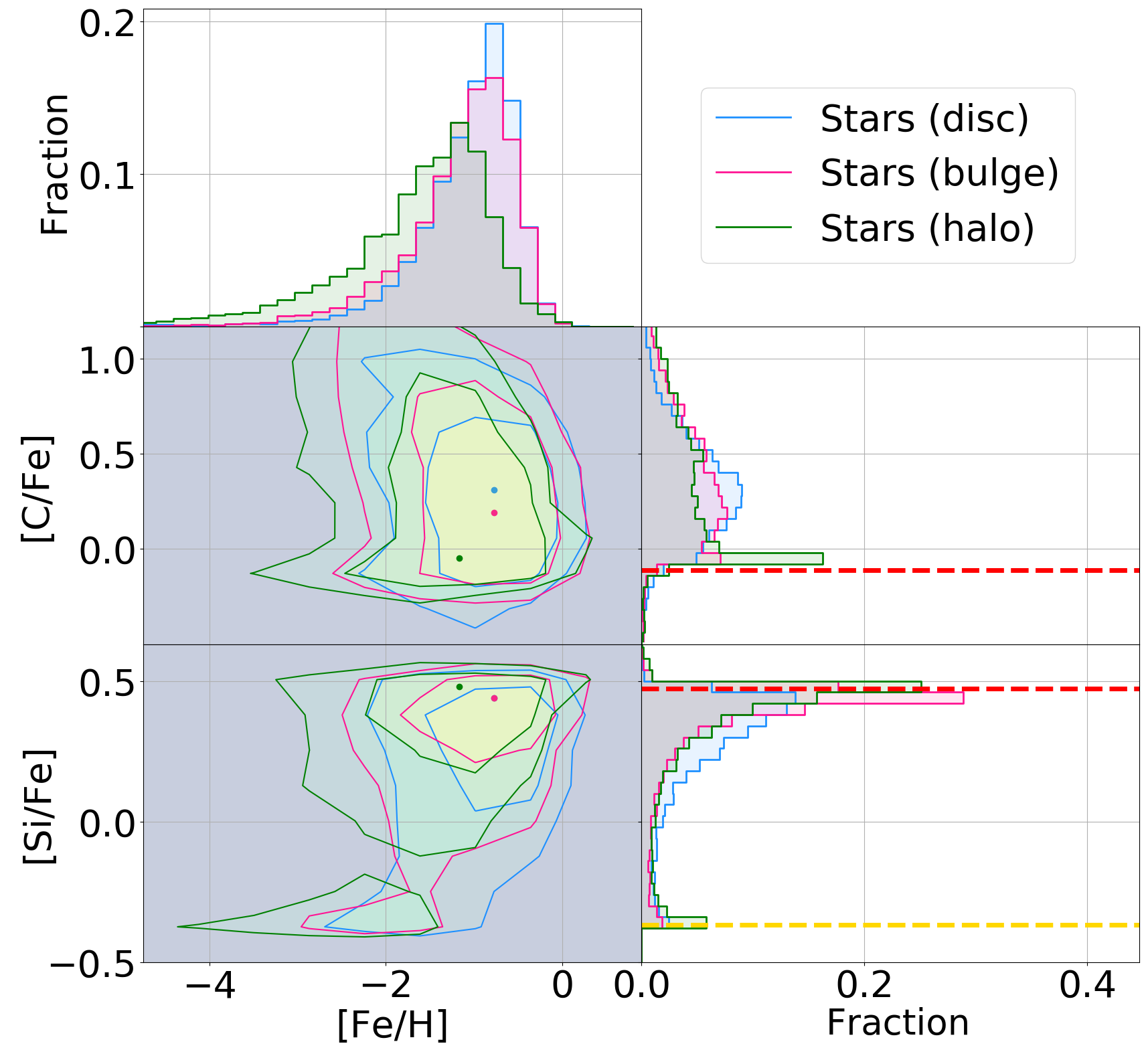}
\caption{\emph{Upper left panel:} relative abundance distributions for [Fe/H] for the three stellar components. \emph{Middle and lower right panels:} [C/Fe] and [Si/Fe] distributions, respectively. \emph{Middle and lower left panels:} distributions in the [Fe/H]-[C/Fe] and [Fe/H]-[Si/Fe] planes, respectively. The red dotted lines mark the relative abundance level of yields corresponding to first generation SNe II, while the yellow one shows the level associated to SNe Ia events.}
\label{fig:ab}
\end{figure}

We find that there are clear differences in the chemical abundances of the three stellar components. This can be seen  in Fig. \ref{fig:ab}, where we show the distribution of [Fe/H], [C/Fe] and [Si/Fe]. Note that the main source of Fe are  SNe Ia, while AGB winds and SNe II are the main contributors of  C  and Si, respectively. 

From the [Fe/H] distribution we can infer that the stellar halo is the component with the lowest Fe enrichment: this results from the fact that the stellar halo is composed of the oldest stars, and has a high contribution of stars formed in smaller systems with lower iron levels. The stellar halo also has low C levels, as stars did not have time to be enriched with AGB winds, and is $\alpha$-enriched, due to fast enrichment with SNe II.
% As this abundance ratio is often used as a metallicity tracer, this indicates that the stellar halo has the less contaminated stars.
On the other hand, the disc and the bulge have higher [Fe/H] levels, the former having the highest C enrichment among all stellar components (C is mainly produced by AGB winds, so as the disc forms gradually and at long timescales, it can be enriched via AGB winds). Finally, the bulge has intermediate C abundances, this component has a large contribution of old and intermediate age stars. 

Note that, in these plots, the red dotted lines indicate the relative abundance levels corresponding to first generation SNe II yields, so stars nearing this value can be interpreted as being formed from gas that was only contaminated early on (as the metallicity dependance of the SNe II yields changes the relative abundances of the ejecta for subsequent stellar generations).

The plot also shows the Si abundances, that we use as  a representative of the $\alpha$-elements, which are mainly produced by SNe II. The [Si/Fe] distribution is dominated by the first generation SNe II yield relative abundance level and, as the gradual Fe contribution from SNe Ia takes place, [Si/Fe] levels decline to reach their minimum at the yellow-dotted line, which indicates the relative abundance level corresponding to SNe Ia yields. The broader appearance of the disc [Si/Fe] distribution is once again consistent with younger stars being in the disc, formed from gas which has received more SNe Ia enrichment. The  stellar halo [Si/Fe] displays a peak towards the yellow dotted line, which indicates stars that have formed from gas particles which {\it only} received SNe Ia feedback. Finally, the bulge shows an intermediate level of enrichment with $\alpha$-elements.

% Finally, the halo component has a significant fraction of stars corresponding to SNe II first generation [C/Fe] level, indicating a population of old stars.

\begin{figure}[!t]
\centering
\includegraphics[width=0.8\columnwidth]{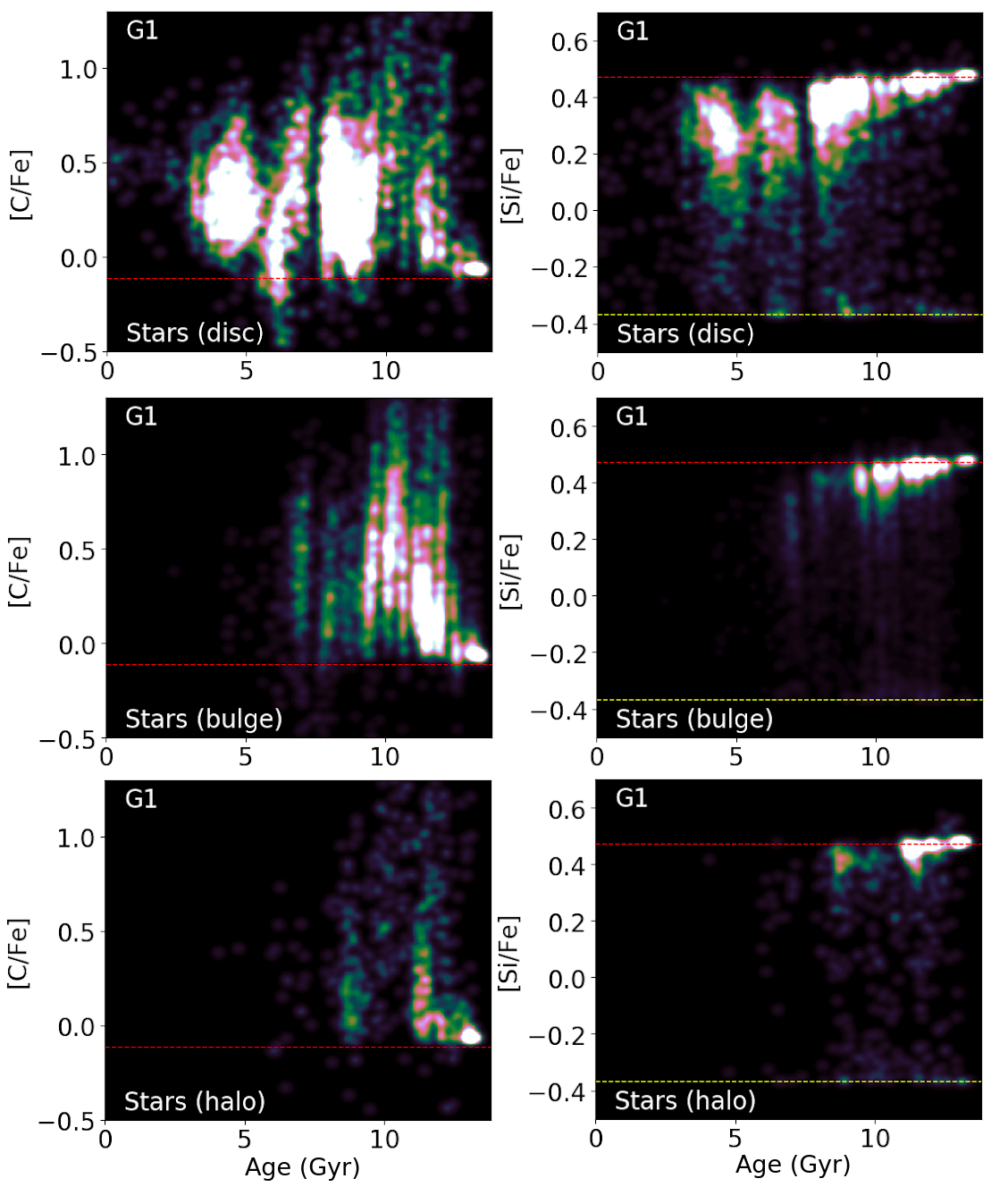}
\caption{\emph{Left panels:} distribution of stars from the three stellar components in the age-[C/Fe] plane. \emph{Right panels:} distributions in the age-[Si/Fe] plane.}
\label{fig:age_ab}
\end{figure}

As explained above, the abundances for the different components are partly the result of the different typical ages of their stars.  
In Fig. \ref{fig:age_ab} we show the stellar distributions in the age-[C/Fe] and the age-[Si/Fe] planes (the age of a star particle is the time elapsed from it's creation to $z = 0$).  This figure confirms that the disc is the component with the youngest stars, the halo has a mostly old stellar population, and the bulge is mainly old but has also contribution of intermediate age stars. From this plot we see that the [C/Fe] peak in the halo distribution shown in Fig. \ref{fig:ab} does correspond to very old stars. It is also evident that the oldest stars show the relative abundance levels corresponding to first generation SNe II, and that the gradual contribution of AGB winds is evidenced in the growth of [C/Fe] levels for younger stars. The same is true for the decline of [Si/Fe] levels owing to Fe enrichment from SNe Ia.

\section{Conclusions}\label{S_con}

We studied the chemical abundance patterns of two Milky Way-sized galaxies in a Local Group simulation, and investigated their relation to the formation history. Both galaxies have 3 well-defined stellar components: the disc, the bulge and the stellar halo.
 We find that $\alpha$-element enrichment occurs early on, due to the fact that SNe II progenitors are short-lived, while Fe and C  enrichment is given in a more gradual fashion, in accordance with the longer time scales of SNe Ia and AGB wind events. This explains the different chemical patterns observed in the disc -- containing the youngest stars ---, the bulge and the stellar halo -- both of which contain mostly old stars.

\begin{acknowledgement}
\texttt{The authors acknowledge support by the Agencia Nacional de Promoción Científica y Tecnológica (ANPCyT, PICT-201-0667).}
\end{acknowledgement}

\bibliographystyle{baaa}
\small
\bibliography{bibliografia}
 
\end{document}